\documentclass[runningheads]{llncs}
\usepackage{booktabs}
\usepackage{xcolor}
\usepackage[T1]{fontenc}
\usepackage{subcaption}
\usepackage{enumitem}

\usepackage{graphicx}
\usepackage{hyperref}
\usepackage[colorinlistoftodos]{todonotes}
\begin{document}
    \title{Improving Low-Resource Retrieval Effectiveness using Zero-Shot Linguistic Similarity Transfer}
\titlerunning{Zero-Shot Linguistic Similarity Transfer}

\author{Andreas Chari\orcidID{0009-0007-9246-8207} \and
Sean MacAvaney\orcidID{0000-0002-8914-2659} \and
Iadh Ounis\orcidID{0000-0003-4701-3223}}

\authorrunning{Chari et al.}

\institute{University of Glasgow, United Kingdom \\
\email{a.chari.1@research.gla.ac.uk, \\ \{Sean.MacAvaney, Iadh.Ounis\}@glasgow.ac.uk}} 

\maketitle  %
\begin{abstract}

Globalisation and colonisation have led the vast majority of the world to use only a fraction of languages, such as English and French, to communicate, excluding many others. This has severely affected the survivability of many now-deemed vulnerable or endangered languages, such as Occitan and Sicilian. These languages often share some characteristics, such as elements of their grammar and lexicon, with other high-resource languages, e.g. French or Italian. They can be clustered into groups of language varieties with various degrees of mutual intelligibility. Current search systems are not usually trained on many of these low-resource varieties, leading search users to express their needs in a high-resource language instead. This problem is further complicated when most information content is expressed in a high-resource language, inhibiting even more retrieval in low-resource languages. We show that current search systems are not robust across language varieties, severely affecting retrieval effectiveness. Therefore, it would be desirable for these systems to leverage the capabilities of neural models to bridge the differences between these varieties. This can allow users to express their needs in their low-resource variety and retrieve the most relevant documents in a high-resource one. To address this, we propose fine-tuning neural rankers on pairs of language varieties, thereby exposing them to their linguistic similarities\footnote{ https://github.com/andreaschari/linguistic-transfer}. We find that this approach improves the performance of the varieties upon which the models were directly trained, thereby regularising these models to generalise and perform better even on unseen language variety pairs. We also explore whether this approach can transfer across language families and observe mixed results that open doors for future research.


\end{abstract}
\section{Introduction}

\begin{figure}[tb]
    \centering
    \includegraphics[width=0.9\linewidth]{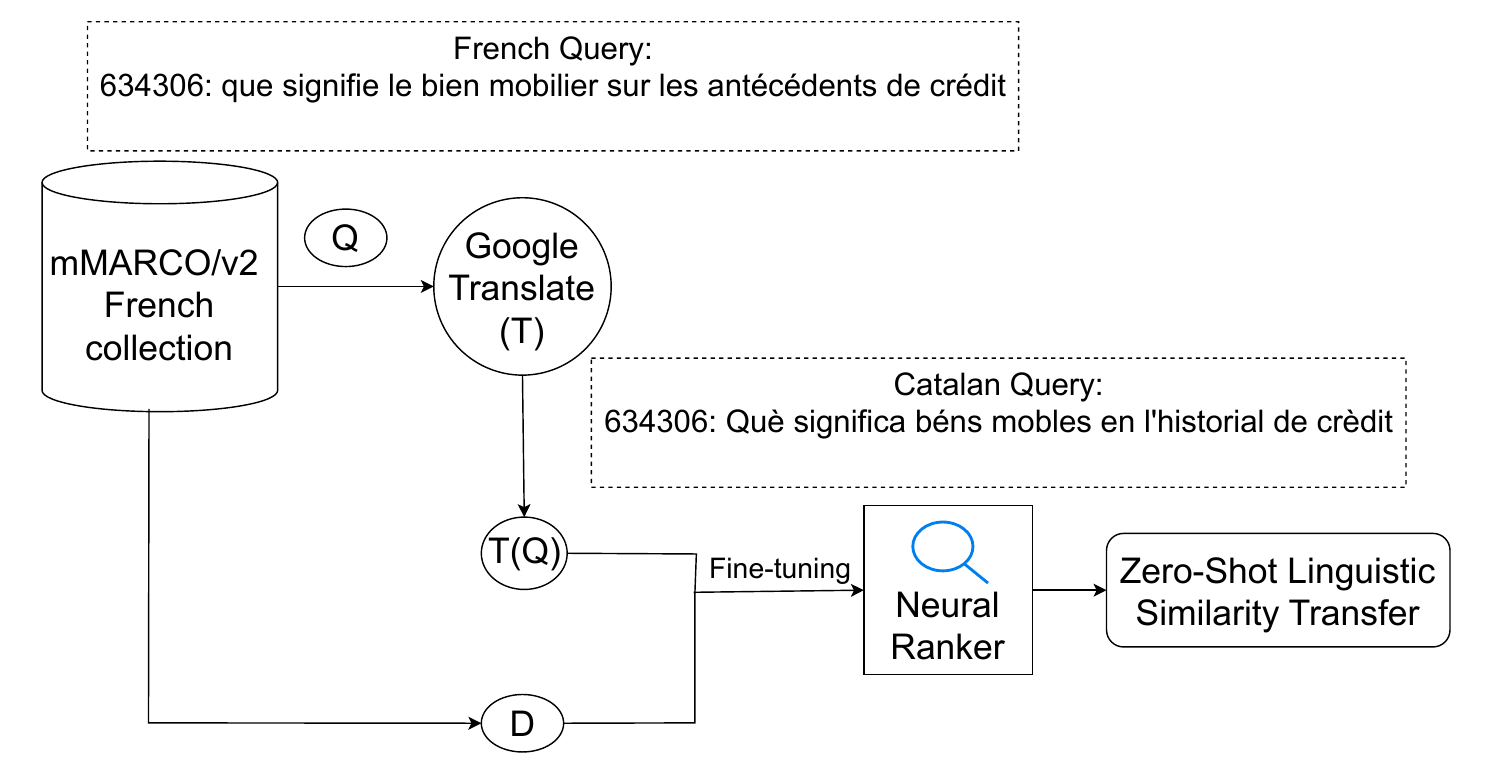}
    \caption{Our Zero-Shot Linguistic Similarity Transfer method leverages a neural ranker fine-tuned on a pair of language varieties, enabling robust retrieval across language varieties even in a zero-shot setting.}
    \label{fig:method}
\end{figure}

The linguistic variability in Information Retrieval (IR) cannot be studied independently of the development of our societies~\cite{DBLP:journals/sigir/Mothe24}. Within this development, globalisation and colonisation were significant determinants in which languages would become the ``lingua franca''\footnote{A \href{https://dictionary.cambridge.org/dictionary/english/lingua-franca}{Lingua Franca} is a language used for communication between groups of people who speak different languages.} and which would not~\cite{doi:https://doi.org/10.1002/9781118332382.ch20,zeng2023english}. In consequence, many of these languages are now considered low-resource due to the lack of digitally recorded linguistic data\footnote{For example, \href{https://fr.wikipedia.org}{French Wikipedia} consists of around 2.6 million articles at the time of writing. In contrast, \href{https://oc.wikipedia.org/}{Occitan Wikipedia} only consists of around 90,000 (3\% of the number of articles).} and even endangered, with UNESCO's World Atlas of Languages listing almost 6000 languages as endangered.\footnote{https://en.wal.unesco.org/}

In sociolinguistics, a variety is a specific form of a language that includes languages and dialects that can be grouped into language clusters~\cite{meecham2001language}.
For instance, Occitan (a language with around 200,000 native speakers), Catalan (a language with 4.1 million native speakers) and French (a language with 74 million native speakers worldwide) are all language varieties with many shared linguistic characteristics and can be clustered as members of the Shifted Western Romance family~\cite{glottolog_shifted_western_romance}. Regarding Occitan, six of its dialects are classified as ``Severely Endangered'' or ``Definitely Endangered'' by UNESCO~\cite{wol-occitan}. In this paper, we aim to leverage the similar linguistic characteristics of these high-resource and low-resource language varieties for ad hoc retrieval.

Ad hoc retrieval is a fundamental task in Information Retrieval (IR), where the goal is to satisfy a user's information need expressed in a query by retrieving some relevant documents from a collection. This task is complicated in the real world when IR search systems must accommodate the cultural and linguistic variability of more than 8 billion people~\cite{un-prospects-2024}. These users will express their information needs in different languages using different grammar, spelling, writing styles, etc. The Glottolog language database~\cite{DBLP:books/daglib/p/Nordhoff12} lists 8,604 languages, of which only a fraction are supported by IR systems.\footnote{For example, the \href{https://lucene.apache.org/core/8_5_1/analyzers-common/index.html}{list of Apache Lucene Analysers} contains Analysers for 38 languages (in addition to an Analyser for ``Indian languages''). \href{https://ir-datasets.com/miracl.html}{MIRACL}, a multilingual retrieval dataset used to evaluate neural rankers, covers only 18 languages.} These systems are not usually trained or have access to information content for low-resource languages, inhibiting the use of low-resource languages for retrieval and often requiring users to search using high-resource ones instead.

Since creating and digitising documents written in low-resource languages is part of longer-term socio-political and interdisciplinary efforts, search researchers can develop systems robust enough to support low-resource language users on the query side. Users will feel empowered to use their language without it being a barrier to searching for information while leveraging the current high-resource document collections. In this work, we leverage Glottolog's language families  and similar to the approach by 
Faisal et al.~\cite{DBLP:conf/acl/FaisalASA0TA24} cluster them into pairs of language varieties. Given that these language varieties share linguistic characteristics and, therefore, information, for the purposes of ad hoc retrieval, queries of similar language varieties should also behave similarly when retrieving relevant documents. 

A common approach to making a neural ranker robust in languages without substantial training data is first to produce automatic translations of the queries and/or documents on these language varieties of a corpus with large quantities of well-annotated data. Then, train the rankers in these languages using a contrastive learning approach. The lack of automatic translation methods for endangered or low-resource languages severely limits this approach. Additionally, in an era where IR research should be more ``Green''~\cite{DBLP:conf/sigir/ScellsZZ22}, repeating this energy-intensive approach for every language we want to support becomes unsustainable and financially impossible to accomplish outside of a small selection of institutions. Therefore, we should leverage the shared linguistic characteristics between low-resource and high-resource language varieties to reduce the training resources needed. We can produce robust and sustainable IR systems if IR research begins to reuse and leverage existing language resources and generalise their gains to other languages with fewer resources.

We start our investigation by using the mMARCO/v2 dataset~\cite{DBLP:journals/corr/abs-2108-13897}, a corpus consisting of automatically translated versions of the MSMARCO passage collection~\cite{DBLP:conf/nips/NguyenRSGTMD16}, and using Glottolog, identify the closest variety for five of the supported languages.\footnote{We focus on five of the thirteen languages since only those had relevant varieties supported by Google Translate.} We then translate the `dev/small' set of those languages to the variety identified and evaluate how robust commonly used lexical and neural IR methods are with queries from pairs of language varieties on the high-resource variety documents. We demonstrate that these widely used retrieval methods cannot generalise to other language varieties. 

Therefore, to overcome this problem, we propose fine-tuning them on two pairs of language varieties: Catalan queries with French documents and Afrikaans queries with Dutch documents. This training will expose the rankers to query-document pairs of different yet similar language varieties. We translate the mMARCO training queries from French to Catalan and Dutch to Afrikaans. Using those translated queries, we fine-tune three models: a multi-vector dense retriever, ColBERT-XM~\cite{DBLP:journals/corr/abs-2402-15059}, a single-vector dense retriever, BGE-M3~\cite{DBLP:journals/corr/abs-2402-03216} and a cross encoder, mT5~\cite{DBLP:conf/naacl/XueCRKASBR21}, to investigate if this supervised training method can expose the models to the shared characteristic of these varieties and close their retrieval effectiveness gap. Our proposed method shown in Figure~\ref{fig:method} allows us to significantly close the gap between the two varieties' effectiveness, leading to performance gains even on the high-resource variety.

In addition, we investigate if we can transfer these gains in performance to language varieties unseen during fine-tuning, which would significantly advance the effectiveness of neural IR methods on low-resource language varieties and reduce the training resources needed. We repeat our initial experiments and observe that this approach does transfer, and similar gains can be observed even in language varieties distinct from either of our training queries' languages, such as Mandarin Chinese. Finally, we investigate whether this approach can be applied across language families. We can observe improvements in the in-domain task and mixed results in the out-of-domain cross-language task. These findings show that this approach does not yet fully transfer across language families as effectively as it does across varieties of the same family. 

Our main contributions can be summarised as follows:

\begin{itemize}
    \item We offer the first to our knowledge evaluation of commonly used retrieval models on similar language varieties.
    \item We propose a zero-shot transfer-based approach that can close the effectiveness gap between language varieties and, in some cases, improve the robustness of both varieties 
    \item We release translated versions of the mMARCO/v2 ``train/judged'' set on Catalan and Afrikaans and ``dev/small'' on six languages alongside model checkpoints for three neural rankers\footnote{https://huggingface.co/datasets/andreaschari/mmarco-lt}.  
\end{itemize}

\section{Related Work}

The recent adoption of pre-trained language models has allowed IR research to exploit their linguistic capabilities for retrieval. While most neural IR research focuses solely on monolingual English IR with models such as ColBERT~\cite{DBLP:conf/sigir/KhattabZ20} or monoT5~\cite{DBLP:conf/emnlp/NogueiraJPL20}, recent years have seen the use of multilingual pre-trained language models such as mDPR \cite{DBLP:conf/nips/AsaiYKH21,DBLP:journals/corr/abs-2408-11942}, XLM-R~\cite{DBLP:conf/acl/ConneauKGCWGGOZ20} and mT5~\cite{DBLP:conf/naacl/XueCRKASBR21} as base models for neural ranking in multiple languages. This has allowed considerable achievements in improving the capabilities of non-English linguistic representation and significant contributions in cross-language and multilingual IR. 

Nair et al.~\cite{DBLP:conf/ecir/NairYLDMMMO22} proposed two methods to generalise ColBERT on a cross-language task: \textit{zero-shot}, where the retriever is solely trained on MS MARCO and relying on the XLM-R encoder for the cross-lingual generalisation and \textit{translate-train} where their ColBERT-X model was trained on pairs of the queries of MSMARCO and automatically translated documents in Chinese, Persian and Russian. Lawrie et al.~\cite{DBLP:conf/ecir/LawrieYOM23} leveraged XLM-R alongside mixed-language batches and translate-train to propose an efficient method for MLIR. From both works, we draw on the compelling performance of methods leveraging translate-train to extend it to a language variety setting where the machine-translated query and document pairs are much closer than a standard cross-language task. A parallel direction to our approach is recent literature leveraging distillation techniques. Yang et al.~\cite{DBLP:conf/ecir/YangLMOM24,DBLP:conf/sigir/YangLM24} recently proposed \textit{Translate-Distill}, where a strong cross-encoder is leveraged as a teacher to distil its knowledge more efficient bi-encoder models in both a CLIR and MLIR setting as an alternative to \textit{translate-train}. Zhuang et al.~\cite{DBLP:conf/sigir/ZhuangSZ23} propose to augment the documents' representations with queries generated using mT5 in languages other than the original document language to mitigate the data scarcity of CLIR. Whereas this approach focuses on enriching the document representation, we focus on more sustainable improvements on the query side of the retrieval pipeline. 

Recent work, such as Chen et al.~\cite{DBLP:journals/corr/abs-2402-03216}, has also been built on top of XLM-R to produce a model that can support retrieval in multiple languages. The authors propose a multi-stage training containing a curated mix of corpora covering almost 200 languages combined with hybrid retrieval and self-knowledge distillation, yielding BGE-M3, a model capable of doing multi-vector, single-vector and sparse retrieval. Similarly, Zhang et al.~\cite{DBLP:journals/corr/abs-2407-19669} proposed the mGTE model which assembles an immense quantity of training data and techniques such as Matryoshka representation learning \cite{DBLP:conf/nips/KusupatiBRWSRHC22} to construct a hybrid encoder and reranker whose performance matches BGE-M3. While both approaches are a helpful resource for encoding non-English text, they both rely on a gigantic amount of training data, most not evenly distributed on all training languages and an unsustainable number of GPUs for training, with BGE-M3 and mGTE both reporting 152 and 64 total GPUs, respectively. 

In direct contrast to these approaches, the recent work by Louis et al.~\cite{DBLP:journals/corr/abs-2402-15059} proposed ColBERT-XM, which leverages a modular approach that is considerably more sustainable and requires fewer resources to train. The approach uses XMOD~\cite{DBLP:conf/naacl/PfeifferGLLC0A22}, a modular transformer, to circumvent the infamous curse of multilinguality~\cite{DBLP:journals/corr/abs-2311-09205}. They propose fine-tuning the retrieval model only in MSMARCO and then zero-shot transferring it to other multilingual tasks leveraging XMOD's adapters. Similarly, MacAvaney et al.~\cite{DBLP:conf/ecir/MacAvaneySG20} also attempted to zero-shot a retrieval system trained on English to non-English collections for MLIR. We build on these works by fine-tuning the retriever on a single cross-variety dataset and zero-shot transferring its knowledge to other language varieties to reduce the training data needed, especially for low-resource varieties without high-quality data. 

Regarding the training data literature, mMARCO~\cite{DBLP:journals/corr/abs-2108-13897} is a popular resource containing automatically translated versions of MSMARCO. CLIRMatrix~\cite{DBLP:conf/emnlp/SunD20} and AfriCLIRMatrix~\cite{DBLP:conf/emnlp/OgundepoZSDL22} were recently proposed, both of which leverage data mined from Wikipedia. In recent years, there has also been a growing interest in evaluating cross-lingual information retrieval. HC4~\cite{DBLP:conf/ecir/LawrieMOY22} and HC3~\cite{DBLP:conf/sigir/LawrieMOYNG23} were recently proposed to benchmark IR methods when dealing with Chinese, Persian, and Russian documents. Similarly, the TREC NeuCLIR track~\cite{DBLP:journals/corr/abs-2304-12367,DBLP:journals/corr/abs-2404-08071} released both NeuCLIR 1, a collection of Persian, Chinese and Russian Common Crawl News documents as well as neuMARCO, a cross-language machine-translated version of MSMARCO to study the effect of neural approaches to CLIR. Both of these collections are now standard benchmarks for evaluating the cross-language capabilities of IR methods. Mr. TYDI~\cite{DBLP:journals/corr/abs-2108-08787} and MIRACL~\cite{DBLP:journals/corr/abs-2210-09984}, both made from text from native speakers, were introduced recently as benchmarks for MLIR. While all of these are excellent resources, none currently attempts to benchmark the robustness of IR methods when dealing with similar language varieties, opting either for mono-lingual evaluation or cross-lingual evaluation with language from entirely different language families and different linguistic characteristics. 

A gap in the current literature is the omission of benchmarks that can assess if and how robust retrieval methods are when dealing with text in distinct \textit{yet similar} languages. While the recent work by Chari et al.~\cite{DBLP:conf/sigir/ChariMO23} has investigated the robustness of various neural retrievers on text expressed with regional spelling conventions, their work focuses solely on American and British English. Similarly, the recent work of Cahyawijaya et al.~\cite{cahyawijaya2024thankyoustingraymultilingual} investigated the robustness of LLMs on cross-lingual sense disambiguation on a word level and found them biased towards high-resource languages. We aim to fill this gap by proposing the first to our knowledge evaluation of the robustness of retrieval methods on similar language varieties at a query and document level.

\section{Methodology}
\vspace{-0.1cm}
Our Zero-Shot Linguistic Similarity Transfer training method works as follows: We take a training query set (Q) and automatically translate it to a similar language variety using a translator (T) to produce a translated version of the training queries, T(Q). We fine-tune a neural ranking method (F) on T(Q) with the original training set's respective documents (D) to expose the model to the similar characteristics of the language varieties of T(Q) and (D).

Our training method leverages the linguistic similarity between two language varieties to expose the neural ranker to similar albeit different representations of information. The rankers are accordingly trained in pairs of low-resource variety queries and high-resource variety documents. This, in turn, will act as a regulariser that will tune the model to identify the common patterns across language varieties. This will ensure the rankers are more effective in scenarios where the user uses a low-resource or endangered language variety and yields relevant documents in a related high-resource variety. We then zero-shot transfer the fine-tuned ranker to other language varieties unseen during training. 

\section{Experimental Setup}
For our investigation into the capabilities of retrieval methods in language varieties, we answer the following research questions:

\begin{itemize}[label={}]
    \item\textbf{RQ1:} How robust are commonly used retrieval methods across language varieties?
    \item \textbf{RQ2:} What is the impact of fine-tuning neural retrieval methods on pairs of language varieties?
    \item \textbf{RQ3:} Does supervised fine-tuning on pairs of language varieties transfer to unseen language varieties?
    \item \textbf{RQ4:} Does supervised fine-tuning on language varieties transfer across language families?
\end{itemize}

For RQ1, we first used the Glottolog language database to identify the language variety closest to the languages of the mMARCO/v2 passage collections supported by Google Translate. Following, we use Google Translate to translate the dev/small set of each mMARCO passage collection in the languages shown in Table~\ref{tab:rq1-languages}. These combinations of a single variety passage collection and two language variety queries make up our benchmark for the robustness of retrieval methods to language varieties.

\begin{table}[tb]
\centering
\caption{Languages varieties for mMARCO/v2 queries alongside their language code.}
\resizebox{0.65\columnwidth}{!}{%
\begin{tabular}{@{}lr@{}}
\toprule
Original language & Language(s) Translated \\ \midrule
French (\textcolor[HTML]{de8f05}{FR})            & Catalan (\textcolor[HTML]{0173b2}{CA}) \& Occitan (\textcolor[HTML]{029e73}{OC})\\
Mandarin Chinese (\textcolor[HTML]{ece133}{ZH})  & Cantonese (\textcolor[HTML]{949494}{YUE})          \\
Dutch (\textcolor[HTML]{d55e00}{NL})             & Afrikaans (\textcolor[HTML]{cc78bc}{AF})         \\
Italian (\textcolor[HTML]{fbafe4}{IT})           & Sicilian (\textcolor[HTML]{ca9161}{SCN})          \\
Indonesian (\textcolor[HTML]{0173b2}{ID})        & Malay (\textcolor[HTML]{56b4e9}{MS})             \\ \bottomrule
\end{tabular}
}
\label{tab:rq1-languages}
\end{table}

For RQ2, we translated the train/judged queries of the French mMARCO/v2 collection into Catalan using Google Translate. We chose French due to the similar linguistic characteristics with Catalan. Both are Shifted Western Romance language varieties, and Catalan shares more linguistic similarities with Northwestern Shifted Romance varieties such as French than other languages such as Italian or Spanish. While there is no document collection for Occitan —the closest language to Catalan— we opted to use the French collection, given their similarity. We repeat the same process for the Dutch queries and translate them to Afrikaans since it is a daughter language of Dutch, and both are varieties of Global Dutch~\cite{glottolog_gdutch}. We then used a zero-shot approach to transfer the information learned from Catalan-French and Dutch-Afrikaans pairs to other language varieties and produce neural retrievers capable of yielding robust performance to low-resource languages while eliminating the need to produce training pairs for every language variety. We experiment with lexical models, bi-encoders and cross-encoders to study how robust these systems are across language varieties.

For RQ1, we benchmark the following models: BM25 \cite{DBLP:conf/trec/RobertsonWJHG94}, ColBERT-XM~\cite{DBLP:journals/corr/abs-2402-15059}, BGE-M3~\cite{DBLP:journals/corr/abs-2402-03216} and mT5~\cite{DBLP:conf/naacl/XueCRKASBR21} as a re-ranker. We focus solely on the neural methods for the remaining research questions and ignore BM25.
For retrieval, we use the PyTerrier platform~\cite{DBLP:conf/ictir/MacdonaldT20} and ir-measures~\cite{DBLP:conf/ecir/MacAvaneyMO22a} except for ColBERT-XM, where we rely on the open-source code provided by its authors.\footnote{https://github.com/ant-louis/xm-retrievers} We report MRR@10 and R@1000 for all experiments with mMARCO and neuMARCO and nDCG@20 and R@1000 for neuCLIR. For datasets, we rely on the ir\_datasets~\cite{DBLP:conf/sigir/MacAvaneyYFDCG21} package, and we mainly use mMARCO/v2~\cite{DBLP:journals/corr/abs-2108-13897} for all research questions, bar RQ4. We additionally use our translated dev/small queries and train/judged queries from mMARCO, which will be released to the community. As for RQ4, we use a cross-lingual setup to test whether our approach transfers across language families and use the neuCLIR/1~\cite{DBLP:journals/corr/abs-2404-08071} and neuMARCO~\cite{DBLP:journals/corr/abs-2304-12367} collections. 

\section{Results and Discussion}

\subsection{RQ1: How robust are commonly used retrieval methods across language varieties?}

\begin{figure}[tb]
    \centering
    \begin{subfigure}{\textwidth}
        \centering
        \includegraphics[width=\linewidth]{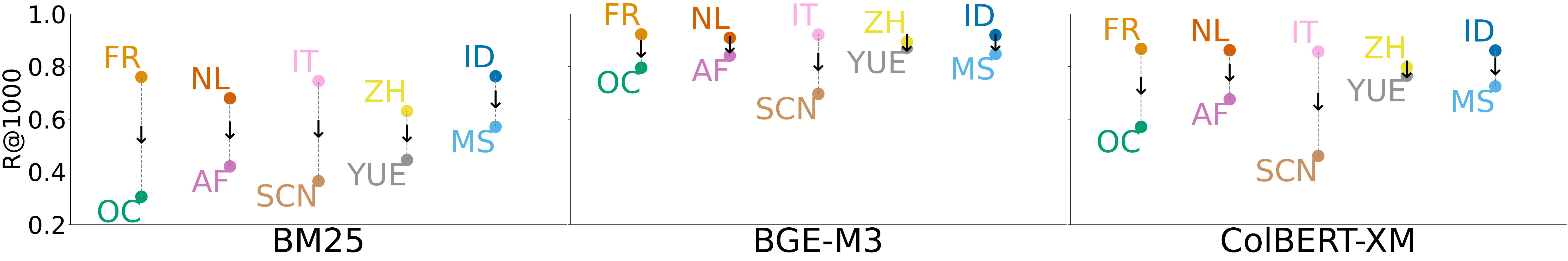}
    \end{subfigure}
    \begin{subfigure}{\textwidth}
        \centering
        \includegraphics[width=\linewidth]{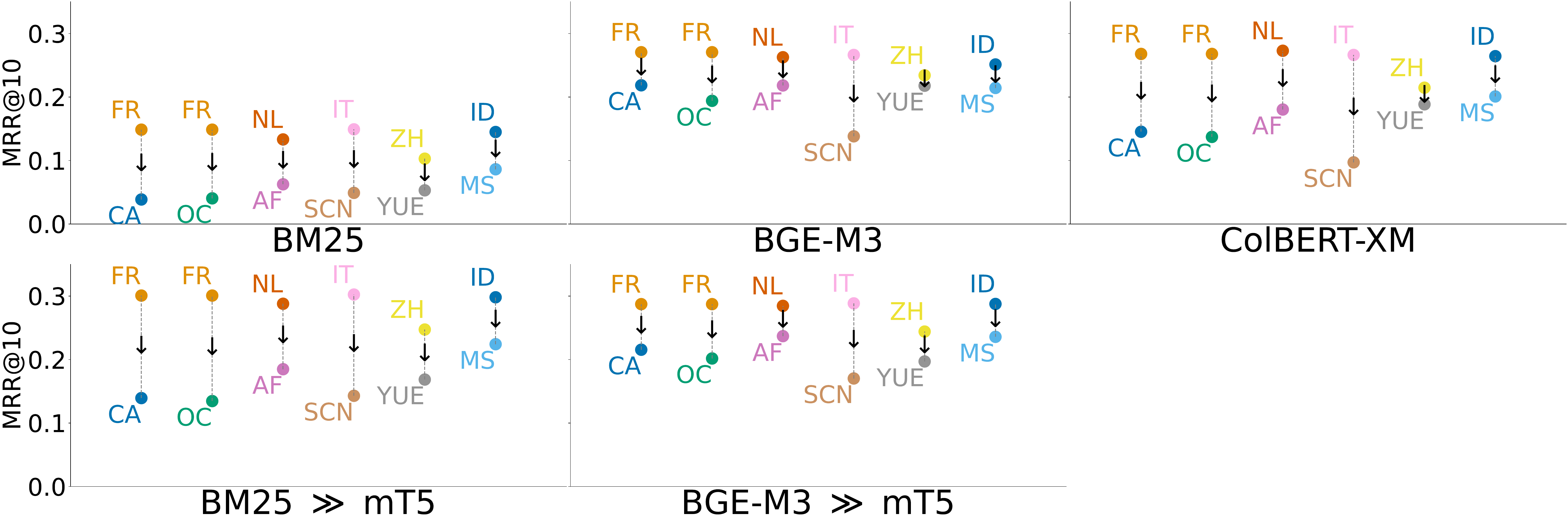}
    \end{subfigure}
    \caption{Retrieval performance of the mMARCO original and translated queries on the documents of the original language variety. All models show significant drops in performance across language varieties. }
    \label{fig:rq1}
\end{figure}

Figure~\ref{fig:rq1} presents our findings for RQ1. We can see significant performance drops for all language varieties in all models. BM25, in particular, sees the most substantial drops in effectiveness, a reasonable consequence of the limitations of lexical retrieval methods. On the other hand, BGE-M3 is the most robust of the first-stage baselines recall-wise. In addition, it shows the smallest drop in performance in the Cantonese vs Mandarin comparison, which is sensible given the sheer volume of Chinese training data the model has seen during its multi-stage training. All models struggle the most with Italian vs Sicilian queries, which is an intriguing observation given the similarities between these varieties. 

A noteworthy observation is that BGE-M3 tends to be more robust than ColBERT-XM in all language varieties. This indicates that the commonly cited argument that multi-vector bi-encoders are more robust than single-vector bi-encoders is not valid for all tasks. However, this might not be specifically due to the model architecture, as BGE-M3 was trained on several more multilingual datasets than ColBERT-XM. Another consistent observation across all baselines is that they seem more robust when dealing with Mandarin vs Cantonese and Indonesian vs Malay queries than the Indo-European languages. In conclusion, we can answer RQ1 that commonly used retrieval methods are not robust across language varieties and show consistent drops in retrieval effectiveness.

\subsection{RQ2: What is the impact of fine-tuning neural retrieval methods on pairs of language varieties?}

\begin{figure}
\centering
    \begin{subfigure}{0.5\textwidth}
        \centering
        \includegraphics[width=\linewidth]{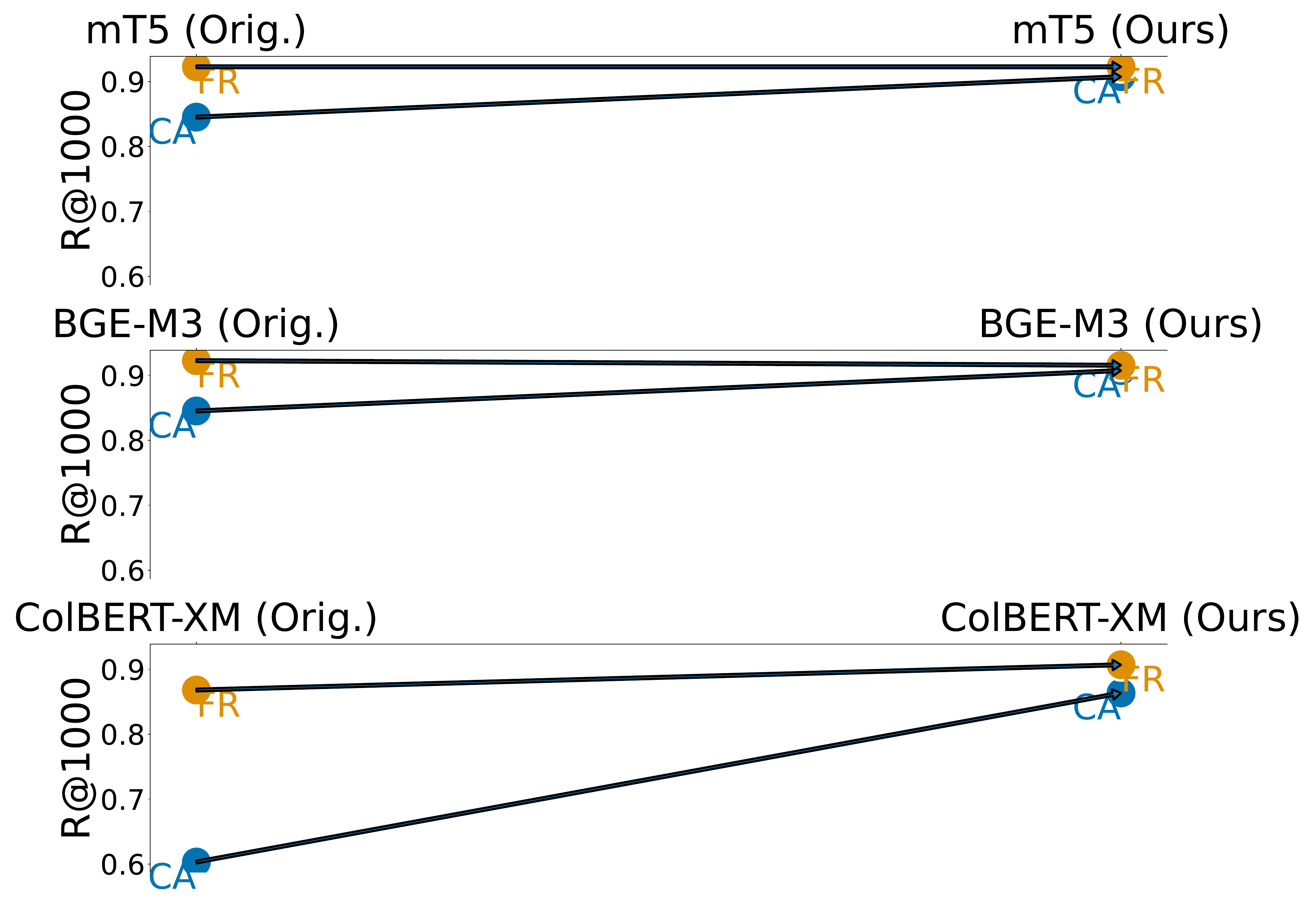}
    \end{subfigure}%
    \begin{subfigure}{0.5\textwidth}
        \centering
        \includegraphics[width=\linewidth]{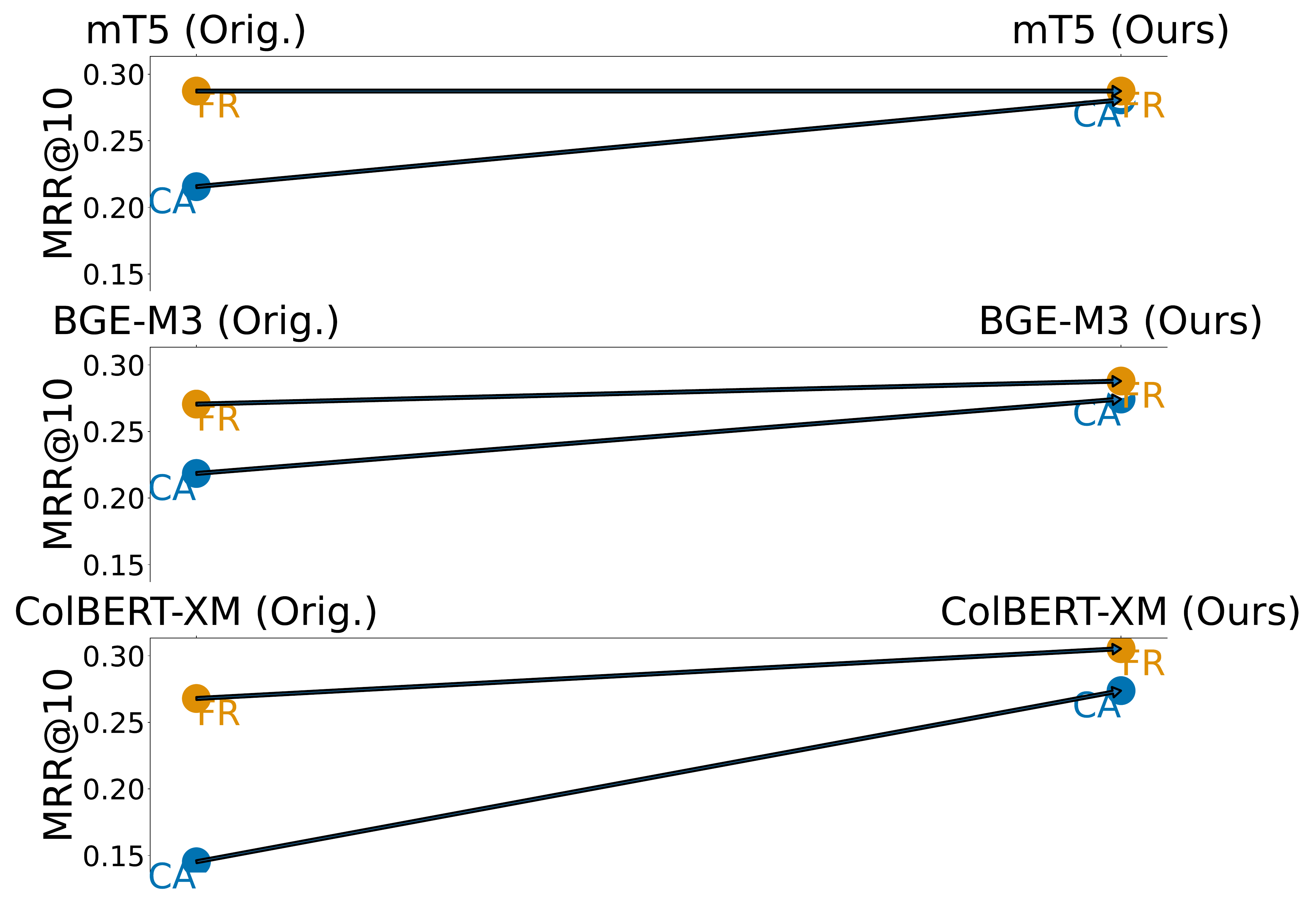}
    \end{subfigure}
     \caption{Performance comparison of baselines \textbf{(Orig.)} and our fine-tuned models \textbf{(Ours)} on the mMARCO French and Catalan queries on the French documents. mT5 recall scores were omitted as they are the same as BGE-M3.}
    \label{fig:rq2-fr}
\end{figure}

Figure~\ref{fig:rq2-fr} presents our findings for RQ2. For all methods, fine-tuning Catalan queries and French documents improves the models' performance in Catalan, with BGE-M3 outperforming the multi-vector ColBERT-XM. Additionally, we observe gains in the retrieval effectiveness of the French queries for the bi-encoders. On the other hand, mT5 remains consistent for French while seeing significant gains in MRR performance for Catalan. 

Overall, ColBERT-XM seems to be the model benefiting the most from this fine-tuning. To answer RQ2, fine-tuning neural retrieval models on pairs of language varieties improves the models' robustness on the low-resource variety whilst occasionally leading to performance gains in the high-resource variety.\footnote{Due to space constraints, we only provide the results of the models trained on the Catalan-French pairs for this and the remaining RQs. They are presented as \textbf{[model-name] (Ours)}. The Dutch-Afrikaans results exhibit similar trends to those of Catalan-French, and we will release them online in supplementary material.}

\subsection{RQ3: Does supervised fine-tuning on language varieties transfer to unseen ones?}

\begin{figure}[tb]
    \centering
    \includegraphics[width=\linewidth]
    {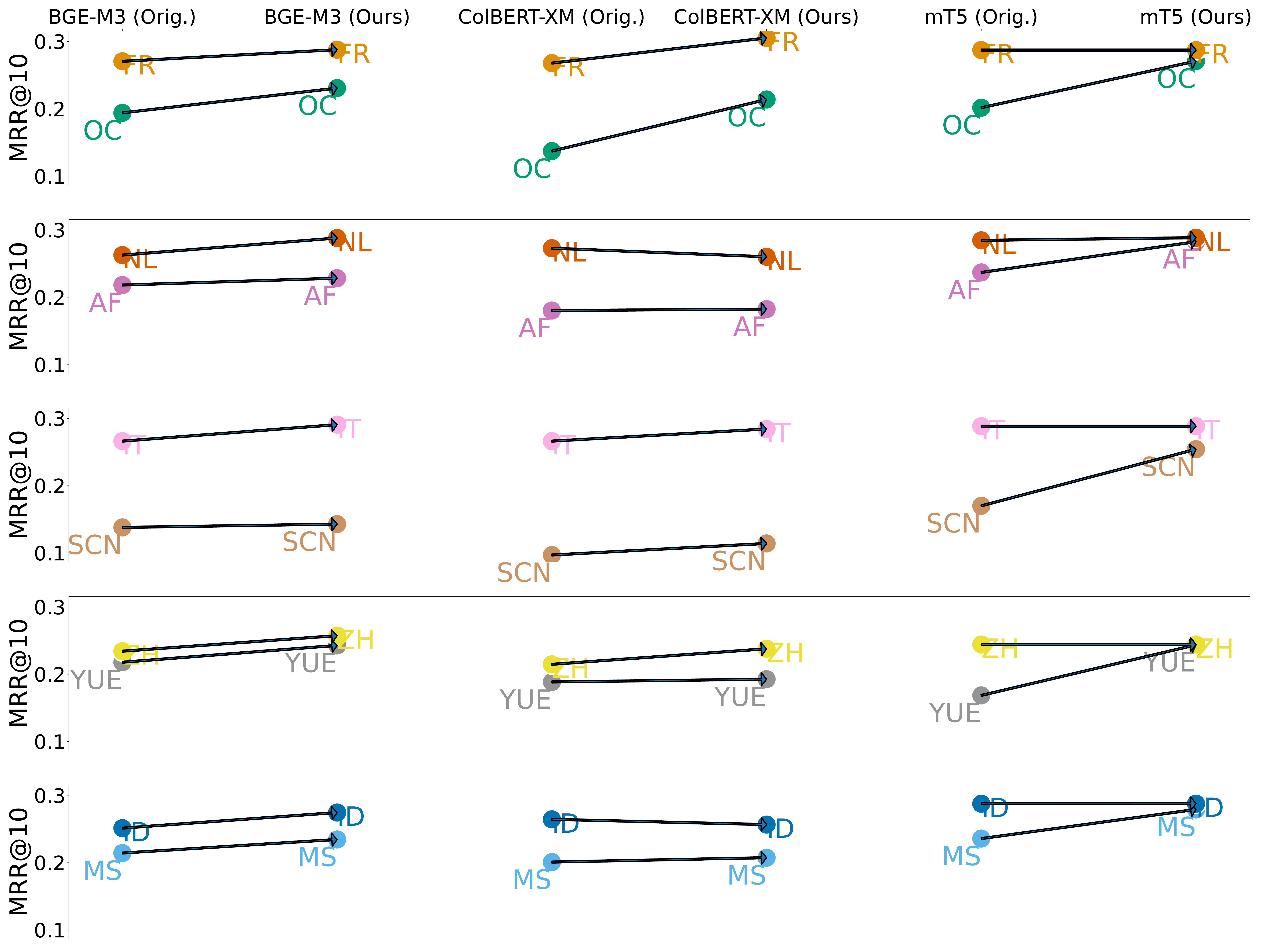}
        \caption{MRR@10 of baselines \textbf{(Orig.)} and our models \textbf{(Ours)} on the unseen mMARCO pairs. \textbf{``mT5''} shows the BGE-M3 results re-ranked with mT5.}
    \label{fig:rq3-mrr-fr}
\end{figure}

\begin{figure}[tb]
    \centering
    \includegraphics[width=\linewidth]{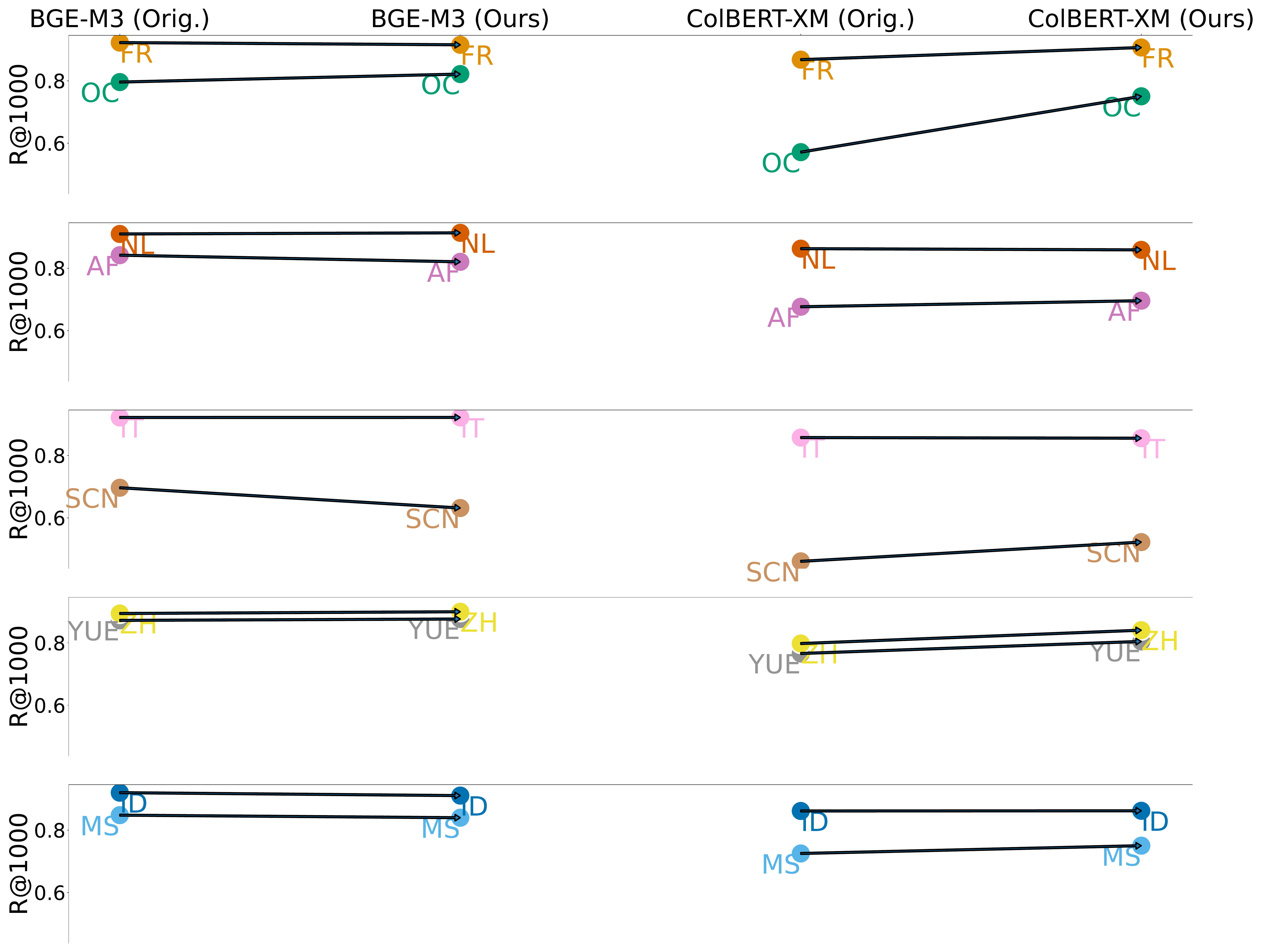}
    \caption{Recall@1000 of baselines \textbf{(Orig.)} and our models \textbf{(Ours)} on the unseen mMARCO pairs. mT5 scores were omitted as they are the same as BGE-M3.}
    \label{fig:rq3-recall-fr}
\end{figure}

Figures \ref{fig:rq3-mrr-fr}~and~\ref{fig:rq3-recall-fr} presents our findings for RQ3. While Recall-wise, BGE-M3's performance is slightly reduced for some languages, we see consistent improvements in MRR for all zero-shot language varieties. Looking at ColBERT-XM, we see consistent gains in both metrics for most language varieties except for Dutch, where the MRR performance slightly decreases. This fine-tuning approach seems to have no major improvements or deterioration in the bi-encoders' performance in the Afrikaans variety. Interestingly, the models seem to have difficulty transferring the Catalan-French fine-tuning to Sicilian, with small gains even though the Italian performance has improved. The most surprising result is that we see the largest transfer gains in the Cantonese-Mandarin pair, varieties with no linguistic connection to either French or Catalan.

Regarding mT5, we observe that the performance of the high-resource variety remains constant for all language pairs. In Addition, we also observe gains in the low-resource varieties with similar trends as the bi-encoders. To answer RQ3, supervised fine-tuning on a pair of language varieties transfers to unseen ones and yields performance improvements on both language varieties.

\subsection{RQ4: Does supervised fine-tuning on language varieties transfer across language families?}

\begin{table}[tb]
\centering
\caption{\textbf{cross-language} evaluation of baselines \textbf{(Orig.)} and our fine-tuned models \textbf{(Ours)} on neuMARCO and neuCLIR.}
\resizebox{0.85\columnwidth}{!}{%
    \begin{tabular}{@{}lccc|ccc@{}}
    \toprule \multicolumn{7}{c}{neuMARCO} \\
    \midrule & \multicolumn{3}{c}{MRR@10} & \multicolumn{3}{c}{R@1000}\\ 
        \midrule
    Model &
      \multicolumn{1}{c}{Farsi} &
      \multicolumn{1}{c}{Russian} &
      \multicolumn{1}{c|}{Chinese} &
      \multicolumn{1}{c}{Farsi} &
      \multicolumn{1}{c}{Russian} &
      \multicolumn{1}{c}{Chinese} \\ \midrule
    BGE-M3 (Orig.) & 0.2143 & 0.2409 & 0.2036 & 0.8721  & 0.8999  & \textbf{0.8314} \\
    BGE-M3 (Ours) & \textbf{0.2382} & \textbf{0.2655} & \textbf{0.2244} & \textbf{0.8779}  & \textbf{0.9010}  & 0.8312 \\ \midrule
    ColBERT-XM (Orig.) & 0.1699 & 0.2134 & 0.1670 & 0.6194  & 0.7341  & 0.5874\\
    ColBERT-XM (Ours) & \textbf{0.1861} & \textbf{0.2239} & \textbf{0.1782} & \textbf{0.7230}  & \textbf{0.7956}  & \textbf{0.6868}\\ \midrule
    BGE-M3 \textgreater{}\textgreater mT5 (Orig.) & 0.2185 & 0.2557 & 0.2076 & 0.8721  & 0.8999  & 0.8314\\
    BGE-M3 \textgreater{}\textgreater mT5 (Ours) & \textbf{0.2855}  &  0.2557 & 0.2076 &  0.8721 &  0.8999 &  0.8314\\
    \midrule
    \multicolumn{7}{c}{neuCLIR} \\
    \midrule & \multicolumn{3}{c}{nDCG@20} & \multicolumn{3}{c}{R@1000}\\ \midrule Model &
      \multicolumn{1}{c}{Farsi} &
      \multicolumn{1}{c}{Russian} &
      \multicolumn{1}{c|}{Chinese} &
      \multicolumn{1}{c}{Farsi} &
      \multicolumn{1}{c}{Russian} &
      \multicolumn{1}{c}{Chinese} \\ \midrule 
    BGE-M3 (Orig.) & \textbf{0.4586}    & \textbf{0.4360}    & \textbf{0.3648} & \textbf{0.6862}  & \textbf{0.6884}  & \textbf{0.6405}\\
    BGE-M3 (Ours)  & 0.3919    & 0.4190    & 0.3360 & 0.5631  & 0.6504  & 0.6059\\ \midrule
    ColBERT-XM (Orig.) & 0.2311    & 0.3200    & 0.2515 & 0.3728  & 0.4917  & 0.4424\\
    ColBERT-XM (Ours) & \textbf{0.2833}    & \textbf{0.3740}    & \textbf{0.2936} & \textbf{0.4097}  & \textbf{0.5831}  & \textbf{0.5094}\\ \midrule
    BGE-M3 \textgreater{}\textgreater mT5 (Orig.) & 0.3551    & 0.4029    & 0.3751 & 0.6862  & 0.6884  & 0.6405\\
    BGE-M3 \textgreater{}\textgreater mT5 (Ours) & 0.3551 & 0.4029 & 0.3751 & 0.6862  & 0.6884  & 0.6405 \\ 
    \bottomrule
    \end{tabular}
}
\label{tab:rq4}
\end{table}

Table~\ref{tab:rq4} shows our experiment results for in- and out-of-domain cross-language evaluation. For the in-domain evaluation, we observe slight performance improvements for the bi-encoders and, curiously, no performance change for the mT5 re-ranker except for Farsi. For the out-of-domain evaluation, we observed contradictory results. The neuCLIR effectiveness of BGE-M3 is severely reduced for all three languages, while for ColBERT-XM, we observe effectiveness gains for all languages. On the other hand, we observe no change in the mT5 re-ranker's retrieval effectiveness. This, alongside our findings for RQ3, indicates that this fine-tuning method only improves mT5's performance when dealing with low-variety languages. 

This inconsistency between in- and out-of-domain evaluation can be caused by the neuCLIR documents being from crawled web pages in contrast to the short web passages of mMARCO. The improvements of ColBERT-XM's effectiveness can also be seen in both in-domain cross-language and the zero-shot cross-variety evaluation for RQ3, which might indicate how beneficial its modular adapters are to generalise on newer languages when fine-tuned to multilingual data. To conclude, we observe mixed results when transferring our method across distinct language families. This indicates the need for further research before our approach can consistently transfer across language families.

\section{Conclusions}

In this paper, we evaluated the performance of commonly used retrieval methods across similar language varieties. We showed that these methods are not robust enough to deal with the differences across language varieties, leading to significant drops in retrieval effectiveness. We showed that fine-tuning these methods on pairs of language varieties, such as Catalan-French and Afrikaans-Dutch, can improve their effectiveness on the trained pairs. We demonstrated how our approach can then be zero-shot transferred to other language varieties unseen during training, even with major linguistic differences such as Mandarin or Indonesian, leading to gains in effectiveness in both high-resource and low-resource language varieties. We additionally found that this approach shows mixed results across distinct language families, opening doors for further research. We hope this paper sparks a paradigm shift in IR research by showing how existing linguistic resources can be leveraged to develop search systems for neglected or under-researched languages.

\subsubsection{Conflict of Interest:}
The authors have no competing interests to declare that are relevant to the content of this article.

\bibliographystyle{splncs04}
\bibliography{bibliography}

\end{document}